# Strong modulation of optical properties in black phosphorus through strain-engineered rippling

*Jorge Quereda[1], Pablo San-Jose[2,]\*, Vincenzo Parente[3], Luis Vaquero-Garzon[3], Aday Molina-Mendoza[1,3], Nicolás Agraït[1,3,4], Gabino Rubio-Bollinger[1,4], Francisco Guinea[3], Rafael Roldán[2,3]\*, Andres Castellanos-Gomez[3]\**

[1] Dpto. de Física de la Materia Condensada, Universidad Autónoma de Madrid, 28049 Madrid, Spain.

[2] Instituto de Ciencia de Materiales de Madrid, CSIC, Sor Juana Ines de la Cruz 3, 28049 Madrid, Spain.

[3] Instituto Madrileño de Estudios Avanzados en Nanociencia (IMDEA-nanociencia), Campus de Cantoblanco, E-18049 Madrid, Spain.

[4] Condensed Matter Physics Center (IFIMAC), Universidad Autónoma de Madrid, E-28049 Madrid, Spain.

(P.S-J.) pablo.sanjose@csic.es , (R.R) rroldan@icmm.csic.es   and (A.C-G.) andres.castellanos@imdea.org

ABSTRACT

Controlling the bandgap through local-strain engineering is an exciting avenue for tailoring optoelectronic materials. Two-dimensional crystals are particularly suited for this purpose because they can withstand unprecedented non-homogeneous deformations before rupture: one can literally bend them and fold them up almost like a piece of paper. Here, we study multi-layer black phosphorus sheets subjected to periodic stress to modulate their optoelectronic properties. We find a remarkable shift of the optical absorption band-edge of up to ~0.7 eV between the regions under tensile and compressive stress, greatly exceeding the strain tunability reported for transition metal dichalcogenides. This observation is supported by theoretical models which also predict that this periodic stress modulation can yield to quantum confinement of carriers at low temperatures. The possibility of generating large strain-induced variations in the local density of charge carriers opens the door for a variety of applications including photovoltaics, quantum optics and two-dimensional optoelectronic devices.

TEXT.

The recent isolation of black phosphorus has unleashed the interest of the community working on 2D materials because of its interesting electronic and optical properties: narrow intrinsic gap, ambipolar field effect and high carrier mobility.[1–12] Black phosphorus is composed of phosphorus atoms held together by strong bonds forming layers that interact through weak van der Waals forces holding the layers stacked on top of each other. This structure, without surface dangling bonds, allows black phosphorus susceptible to withstand very large localized deformations without breaking (similarly to graphene and $MoS_2$).[13–15] Its outstanding mechanical resilience makes black phosphorus a prospective candidate for strain engineering, i.e. the modification of a material's optical/electrical properties by means of mechanical stress.[16] This is in contrast to conventional 3D semiconductors that tend to break for moderate deformations. Very recent theoretical works explore the effect of strain on the band structure and optical properties of black phosphorus, predicting an even stronger response than in other 2D semiconductors such as transition metal dichalcogenides. Most of the reported works, however, are limited to theoretical studies[13,14,17–19] dealing exclusively with uniform strain, while the role of non-uniform strain remains poorly-understood.

Here we explore the effect of periodic strain profiles to modulate the electronic and optical properties of black phosphorus by combining hyperspectral imaging spectroscopy experiments and tight-binding calculations. We spatially probe the shift of the absorption edge in black phosphorus flakes subjected to a periodic strain profile of alternating compressive and tensile stress with periods of ~1 μm. We find that the optical absorption edge shifts towards higher energy at positions with tensile stress while compressive stress produces a shift towards lower energies. The magnitude of the observed shifts, around +10% (tensile) to -30% (compressive) of the unstrained sample gap, greatly exceeds typical values measured in transition metal dichalcogenides.[20–26] The results of our tight-binding simulations are compatible with these observations for strains below ± 5%. Most saliently, the observed spatial modulation of the absorption edge suggests the possibility of strain-induced confinement of carriers along the ripple valleys, where the local electronic gap is smallest. Such confinement is indeed confirmed by our numerical simulations of the system's local density of states. Q*uantum wires* with discrete one-dimensional subbands are theoretically expected to develop in the presence of a non-uniform strain in the sample. These can have a strong impact in photovoltaics, as it is predicted that a spatial modulation of the band gap could be exploited to guide the photogenerated excitons towards the positions with the smaller band gap, thus facilitating the collection of photocurrent.[23,27,28]

Black phosphorus samples have been fabricated by mechanical exfoliation of bulk black phosphorus (Smart Elements) with Nitto tape (SPV 224). All the measurements have been carried out in less than 20 minutes after the black phosphorus exfoliation to minimize its exposure to ambient conditions as it has been shown that black phosphorus degradation is negligible at such short time scales.[2,29] In order to generate a periodic strain profile the exfoliated flakes have been transferred onto a pre-stretched elastomeric substrate (Gel-film®) and subsequently the strain has been suddenly released (see the Materials and Methods section of the main text and the Section 1 of the Supporting Information for more details about the sample fabrication).[23] For large initial pre-stress this process produces large delaminated wrinkles. For moderated initial stress (<30%), on the other hand, the trade-off between the compressive force, the bending rigidity and the adhesion force forms a periodic rippling in the black phosphorus flakes (without delamination). In fact, when a thin film is deposited on a compliant elastic substrate, subjected to moderate compressive uniaxial strain, the thin film buckles and the substrate deforms coherently, forming surface ripples with regions alternating tensile (on the summits) and compressive stress (on the valleys) on the elastomeric substrate that strains accordingly the adhered thin film. We address the reader to Ref. [30] for more details on the mechanisms involved in the buckling-induced-delamination and buckling-induced-rippling processes. Note that, close to the flake edges one can typically find flat regions that have been delaminated from the substrate during the fabrication process, relaxing the accumulated strain.

Figure 1a shows a transmission mode optical microscopy image of a 10 nm thick (~18 layers) black phosphorus flake on a Gel-film® substrate, which has been subjected to the buckling-induced-rippling procedure described above. The thickness of the flake has been determined from its transmittance, measured at different wavelengths (see the Supporting Information for more details). The ripples have a period of ~1 μm. Note that thinner flakes yield shorter rippling period (the period is predicted to be proportional to the flake thickness [30]) and thus we have limited our study to flakes with thickness >10 nm which provide ripple periods larger than the spatial resolution of the experimental setup. The height of the ripples can be estimated by atomic force microscopy (AFM), see the bottom panel in Figure 1a, although the measured topography can be distorted by the finite size of the AFM tip and the narrow spacing between ripple summits, which makes the valleys look artificially shallower, and by the compliance of the elastomeric substrate. One can also determine the relative orientation of the ripples with respect to the black phosphorus lattice by exploiting the strong linear dichroism of black phosphorus (its optical absorption depends on the relative orientation between the materials lattice and incident linearly polarized light).[31,32] The flake shown in Figure 1 presents the ripples almost aligned parallel to the zigzag direction, with an uncertainty of ±5 degrees (see the Supporting Information for more details on the crystal orientation determination). During this work we found that the ripples tend to be preferentially aligned parallel to the zigzag direction (see Section 5 of the Supporting Information) in agreement with recent theoretical predictions that the zigzag direction is about 4 times stiffer than the armchair direction.[13,14]

The effect of the periodic deformation of the black phosphorus on its optical properties can be spatially resolved by means of a recently developed hyperspectral imaging spectroscopy technique, described in detail in Ref. [33]. Figure 1b shows the absorption spectra acquired at different positions on the black phosphorus flake indicated with colored circles: on three ripple summits, on three ripple valleys and on three flat regions. On the bottom part of the ripples the absorption edge (i.e. a photon energy above which the material's absorption sharply increases due to the excitation of carriers from the valence band to the conduction band) occurs at an energy lower than 1.4 eV. Interestingly, the absorption edge is strongly blue-shifted by ~700 meV on the summits relative to the valleys (see arrow). From Figure 1b, it seems that the valleys are more shifted with respect to the flat regions than the summits. We attribute this to the presence of a global compressive strain because of the fabrication method employed that relies on the uniaxial compression of the black phosphorus flake / PDMS stack. Note that the studied multilayer black phosphorus flakes are expected to have a band gap smaller than 1.4 eV and thus the observed sudden increase in the absorption coefficient is not directly the modulation of the band gap. As explained in the theory discussion below, however, this high energy absorption turns out to follow the same modulation as the gap with strain, so that the absorption edge shifts are an indirect measurement of the gap modulation. The whole optical absorption curve is therefore blue- (red-) shifted at positions under tensile (compressive) strain. This is once more in contrast with other 2D semiconductors where an opposite sign of the strain-induced band gap change has been observed.[20–25]

A deeper insight into the spatial variation of the absorption edge energy can be obtained through iso-absorption maps that represent the energy at which the local absorption at each sample position reaches a certain value. In Figure 1c we present an iso-absorption map that shows the spatial dependence of the absorption edge onset across the rippled black phosphorus sample (see the Supporting Information for more details about the hyperspectral technique and data analysis used to build up this map). The energy at which the black phosphorus sample presents the same absorption value follows the pattern imposed by the periodic ripples that induce localized tensile and compressive strains. A line cut along an area containing 11 periods and a flat region shows that the absorption edge shifts by a +10% on the summits and a –30% on the valleys with respect to the flat region (presumably unstrained), which stems from an equivalent strain-induced modulation of the gap. Note that the exact shift of the absorption edge (in %) depends on the absorption value employed to build up the iso-absorption map (see thee Supporting Information) and thus this method should be used to extract qualitative information about the spatial variation of the absorption edge rather than accurate energy values of the absorption edge which should be extracted from the plots displayed in Figure 1b. Nonetheless, these values are much larger than those observed in $MoS_2$ by photoluminescence measurements (up to 5% band gap change for comparable strain levels).[20–25]

The above experimental results can be understood in the framework of a theoretical tight-binding model, where the electron hopping processes depend on the relative angle and on the relative atomic separation between atoms.[34] This technique is especially appropriate to study the response of finite samples to non-uniform strain profiles, which is the case of interest here. First-principle methods, on the other hand, are not well suited to analyze this problem because of the huge unit cell that must be considered, making the calculation extremely expensive in computational terms. Black phosphorus consists in puckered atomic layers of phosphorus weakly coupled together by van der Waals interaction.[35] We use a tight-binding model that considers ten intra-layer and four inter-layer hopping terms (Fig. 2a) which properly accounts for valence and conduction bands for energies ~ 0.3 eV beyond the gap.[36,37] For a given profile of strain, the atoms modify their positions with respect to the equilibrium configuration, their new relative separation defined as $\vec{R}_{ij} = (\mathbb{I} + \varepsilon)\vec{R}_{ij}^0$ where $\mathbb{I}$ is the identity matrix, $\vec{R}_{ij}^0$ is the vector connecting sites $i$ and $j$ in the undistorted lattice, and the well-known anisotropy of the elastic properties of black phosphorus is included in the strain tensor $\varepsilon$, which takes different forms for strain applied along zigzag or armchair directions (see Methods and Supporting Information for more details). The hopping terms $t_{ij}$ depend on $\vec{R}_{ij}$, so they will be modified by a small-to-moderate ripple strain as [38,39] $t_{ij} = t_{ij}^0 \left(1 - \beta_{ij} \frac{\delta R_{ij}}{R_{ij}^0}\right)$, where $t_{ij}^0$ is the corresponding hopping term in the undistorted lattice, $R_{ij} = |\vec{R}_{ij}|$ is the inter-atomic distance in the strained lattice, $\delta R_{ij} = R_{ij} - R_{ij}^0$ and $\beta_{ij} = d \log t_{ij} / d \log R_{ij}$ is the deformation coupling. In the absence of any experimental or theoretical estimation of the coupling, we use the so-called Wills-Harrison [40] argument to assume $\beta_{ij} \approx 3$ (see Materials and Methods and Supporting Information for more details on the tight-binding model and on the elastic properties of black phosphorus).

The calculated electronic band structure for a monolayer with different values of uniaxial strain is shown in Fig. 2b, with the result that the gap increases (decreases) for tensile (compressive) strain, in agreement with our experimental measurements of Fig. 1. This behavior is generic, and is also obtained for multilayers and bulk black phosphorus. The experimental ~18-layer sample is expected to be accurately modeled by the bulk limit, as all spectral and optical properties quickly converge to said limit above ~12 layers.[41] To connect the spectral properties to the optical properties of strained black phosphorus, we have used the Kubo formula to compute the optical conductivity of both undoped single layer and bulk black phosphorus for different values and directions of uniaxial strain. In agreement with previous results for unstrained black phosphorus,[42,43] the material exhibits a strong linear dichroism, i.e. a strong difference in optical conductivity for incident light polarized along zigzag $\sigma_{ZZ}(\omega)$ and armchair $\sigma_{AC}(\omega)$ directions, normalized to $\sigma_0 = 2e^2/h$ (see Fig. 2c-d). The modulation of gap with strain is directly observed in optical conductivity as a blue (red) shift of the onset at photon energy equal to the gap, above which optical transitions are allowed, although in the bulk limit it lies at energies ~0.3 eV, below our detection threshold. As shown in Figure 2c-d, however, the whole $\sigma_{AC}(\omega)$ curve jointly shifts with strain similarly to the band gap (Figure 2b). The expected change in bulk $\sigma_{AC}(\omega)$ with a $\pm$ 5% strain modulation in the energy window of our experiment is in the ~ 800 meV range, compatible with the observed shift in the optical absorption of Figure 1b.

Having analyzed the electric and optical properties of black phosphorus under homogeneous strain, we now exploit the tight-binding model to study the case of non-uniform strain relevant to our rippled samples. We consider an undulated monolayer with 500 nm period ripples (perpendicular to the armchair direction as in the experiment), subjected to a sinusoidal strain profile as the one sketched in Figure 3a, where maximal tensile and compressive regions occur at the summit and valleys of the ripples, respectively (see Supporting Information for more details on the calculation, and results for the zig-zag case). The local density of states (LDOS) corresponding to such configuration is presented in Figure 3b, which shows how the local gap is maximum at the summits, where the sample is under tension, and minimum at the valleys, for which it is maximally compressed. Such numerical results agree with the experimental optical absorption spectra of Figure 1, and confirm the extraordinary tunable properties of black phosphorus with strain. Notably, moreover, they show that the gap modulation, even for the narrow 500 nm ripples considered in our simulation, gives rise to strong quantum confinement of carriers within ripple valleys at low enough temperatures. This is apparent from the formation of quasi-1D subbands in the LDOS (sharp van-Hove resonances as a function of energy). The spatial width of the lowest mode is a mere ~130-160 nm, with a distance to the next subband exceeding 20 meV even in the multilayer case (see Supporting Information for more details). While our experimental observations are performed at room temperature, and are thereby unable to resolve discrete subbands, they however unambiguously demonstrate that carriers are fully confined along the ripples, since otherwise the observed absorption edge would not be spatially modulated.

It is interesting to summarize our results by comparing the effect of strain in black phosphorus to other two dimensional crystals, like transition metal dichalcogenides.[20–23] Two considerations are in order: first, as we have seen, tensile (compressive) strain opens (closes) the gap in black phosphorus, whereas the opposite trend is observed for transition metal dichalcogenides. Second, the black phosphorus band structure is considerably more sensitive to strain than the electronic structure of transition metal dichalcogenides, spanning a range of around 0.7 eV in band modulation even with moderate strains. This is a consequence of the peculiar puckered lattice structure of black phosphorus layers. The gap in this material is mainly controlled by the ratio between in-plane and out-of-plane hoppings ($t_1^\parallel$ and $t_2^\parallel$ in Fig. 2a, respectively), which are affected very differently by in-plane strains. This is in contrast to the case of transition metal dichalcogenides, whose gap essentially stems from onsite crystal fields on the metal orbitals, and is therefore much less sensitive to strains.

Our results moreover suggest that black phosphorus is an extraordinary candidate to create exciton funnels,[27] but unfortunately our spatially resolved optical spectroscopy technique is not suitable to probe these excitonic effects. The funnel effect arises in the presence of a spatial modulation of the band gap in a reduced spatial region. As we have shown, this modulation is huge in our samples, of up to ~0.7 eV. This is expected to lead to strong confinement of carriers into narrow one-dimensional ~150 nm wide *quantum wires* formed along the valleys of the sample ripples (see Figure 3). If the lifetime of carriers is large enough to permit sufficient exciton drift before recombination, the excitons will be funneled into the quantum wires until the recombination takes place. Funneling has been recently proposed as a powerful strategy to enhance the efficiency of photovoltaic energy harvesting devices by facilitating the collection of photogenerated carriers.[23,27,28] More fundamentally, it has also been discussed as an interesting avenue to create exciton condensates in solid state devices, with trapping potentials ~20 meV that are much stronger than those achieved by alternative implementations based on the AC Stark effect (reaching 5 meV confining potential).[44]

MATERIALS AND METHODS

We prepared black phosphorus nanosheets on elastomeric substrates (Gel-Film® PF 6.5mil ×4 films) by mechanical exfoliation of bulk black phosphorus crystals (Smart Elements) with blue Nitto tape (Nitto Denko Co., SPV 224P). Gel-Film® is a commercially available elastomeric film based on poly-dimethyl siloxane (see the Supplementary Information in Ref. [[23]] for a characterization of the Gel-Film® material). We located few-layer black phosphorus sheets under an optical microscope (Nikon Eclipse LV100) and determined the number of layers by their opacity in transmission mode (see the Supplementary Information for more details on the thickness determination of the flakes).

The topography of the ripples has been characterized by atomic force microscopy (AFM). A *Nanotec Electronica Cervantes AFM* microscope (with standard cantilevers with spring constant of 40 N/m and tip curvature <10 nm) operated in the amplitude modulation mode has been used to study the topography of the ripples.

The hyperspectral imaging setup used here has a spatial resolution of $d = \lambda/2NA$, where $d$ is the smallest resolvable feature $\lambda$ the illumination wavelength and NA the microscope objective numerical aperture (NA = 0.8 in our experimental setup). Which gives a spatial resolution value of ~550 nm at the longer wavelengths employed in this work.

Theoretical modeling of the electronic band structure in the presence of local strain was performed by using a tight-binding Hamiltonian $H = \left( \sum_{\langle\langle i,j \rangle\rangle} t_{ij}^{\parallel} c_i^{\dagger} c_j + \sum_{\langle\langle i,j \rangle\rangle} t_{ij}^{\perp} c_i^{\dagger} c_j \right) + h.c.$ where $c_i^{\dagger} (c_i)$ creates (annihilates) an electron at site $i$. and The intra-layer hopping terms $t_{ij}^{\parallel}$ considered in the model are shown in Fig. 2a with values that have been obtained by fitting to *ab initio* calculations [36,37]: $t_1^{\parallel} = -1.486$ eV, $t_2^{\parallel} = 3.729$ eV, $t_3^{\parallel} = -0.252$ eV, $t_4^{\parallel} = -0.071$ eV, $t_5^{\parallel} = -0.019$ eV, $t_6^{\parallel} = 0.186$ eV, $t_7^{\parallel} = -0.063$ eV, $t_8^{\parallel} = 0.101$ eV, $t_9^{\parallel} = -0.042$ eV and $t_{10}^{\parallel} = 0.073$ eV. Multilayer samples are considered by including four inter-layer hopping terms $t_{ij}^{\perp}$ in the Hamiltonian, with values: $t_1^{\perp} = 0.524$ eV, $t_2^{\perp} = 0.180$ eV, $t_3^{\perp} = -0.123$ eV and $t_4^{\perp} = -0.168$ eV. In the presence of an external strain field, the hopping terms are modified as $t_{ij} = t_{ij}^0 \left( 1 - \beta_{ij} \frac{\delta R_{ij}}{R_{ij}^0} \right)$, where $t_{ij}^0$ and $R_{ij}^0 = |\vec{R}_{ij}^0|$ refer to the hopping and inter-atomic distances in the undistorted lattice, $\delta R_{ij} = R_{ij} - R_{ij}^0$ and $\beta_{ij} = d \log t_{ij} / d \log R_{ij}$ is the deformation coupling [38,39]. The separation between $i$ and $j$ sites in the presence of strain can be obtained as $\vec{R}_{ij} = (\mathbb{I} + \boldsymbol{\varepsilon})\vec{R}_{ij}^0$ where $\mathbb{I}$ is the identity matrix and $\boldsymbol{\varepsilon}$ is the strain tensor. Black phosphorus is known to be an extremely anisotropic crystal, with the in-plane anisotropy of the elastic properties being encoded in the strain tensor $\boldsymbol{\varepsilon}$, defined as $\boldsymbol{\varepsilon}_{ZZ} = \epsilon(x) \begin{pmatrix} 1 & 0 & 0 \\ 0 & -\nu_{yx} & 0 \\ 0 & 0 & 0 \end{pmatrix}$ for uniaxial strain along the zigzag direction, and $\boldsymbol{\varepsilon}_{AC} = \epsilon(y) \begin{pmatrix} -\nu_{xy} & 0 & 0 \\ 0 & 1 & 0 \\ 0 & 0 & -\nu_3 \end{pmatrix}$ along the armchair direction. In the above expressions, $\nu_{xy} \approx 0.2$ and $\nu_{yx} \approx 0.7$ are the corresponding Poisson ratios, which account for the in-plane deformation of the crystal in the perpendicular direction to the applied strain.[13] Whereas an externally applied uniaxial strain along the zigzag direction does not lead to a significant deformation in the perpendicular direction, this is not the case if the strain is applied along the armchair direction, which induces a considerably flattening (widening) of the lattice for tensile (compressive) strain. This effect is accounted for by the Poisson ratio, which (absent quantitative experimental or numerical information) we take as $\nu_3 \approx 0.2$ for single-layer phosphorene [19] and $\nu_3 \approx 0.1$ for bulk black phosphorus, estimated as $\nu_3 = |C_{13}/C_{11}|$ from the elastic constants calculated in Ref. [45]. To compare with our experiments, the modulation of the uniaxial strain is chosen to be defined by a sinusoidal function $\epsilon(y) = \frac{\epsilon_0}{2}\cos(2\pi y/L)$ where $L$ is the period of the undulation and $\epsilon_0 \approx 0.05$ is the estimated strain difference between the summits (which are tensioned with strain $+\epsilon_0/2$) and the valleys (which are compressed with strain $-\epsilon_0/2$). The calculation of band structure, optical conductivity and local density of states based on the above tight-binding model have been performed using the MathQ package, developed by P. San-Jose. Source: http://www.icmm.csic.es/sanjose/MathQ/MathQ.html. (See Supporting Information for more details on the theoretical model).

FIGURES

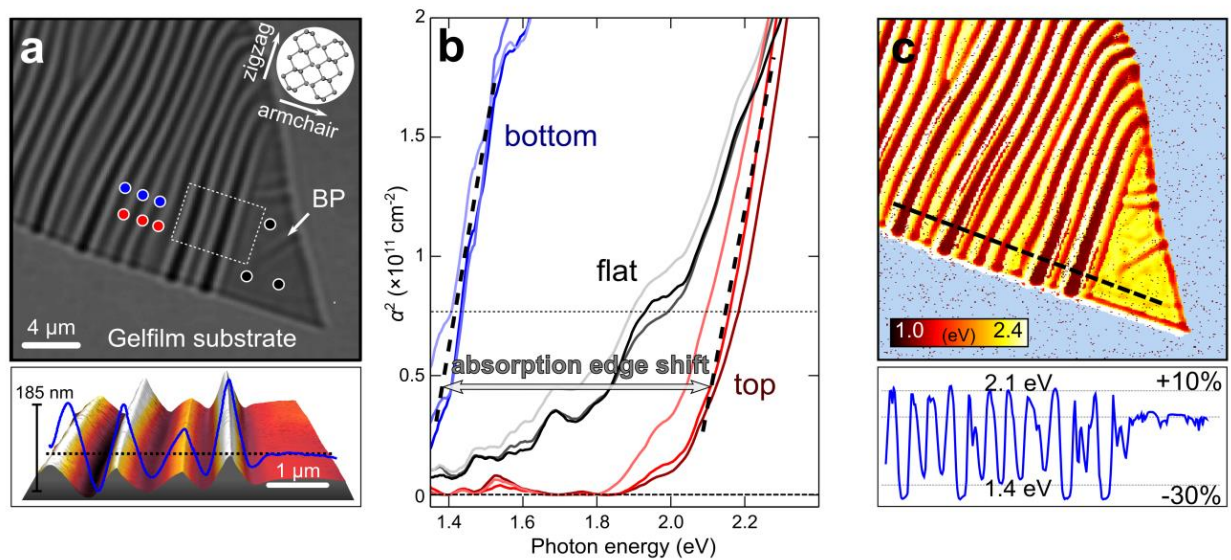

**Figure 1. Experimental realization of periodically strained black phosphorus**. **a** Transmission-mode optical image of a ripples 10 nm thick black phosphorus flake. The inset shows a sketch of the crystal lattice orientation, determined from the measurement of its linear dichroism (see Supporting Information for more details). Below the optical image, an atomic force microscopy topography image acquired on the region highlighted with the dashed rectangle in **a** is shown. **b** Optical absorption spectra acquired on three ripple summits, three valleys and three flat regions, indicated with coloured circles in **a**. **c** Iso-absorption map that represents the energy at which $\alpha^2$ = 7.5·10$^{-11}$ cm$^{-2}$ at each sample location. The map illustrates how the absorption spectra spatially varies due to the periodic compressive and tensile stress induced by the ripples. A linecut along the dashed line is showed below the iso-absorption map.

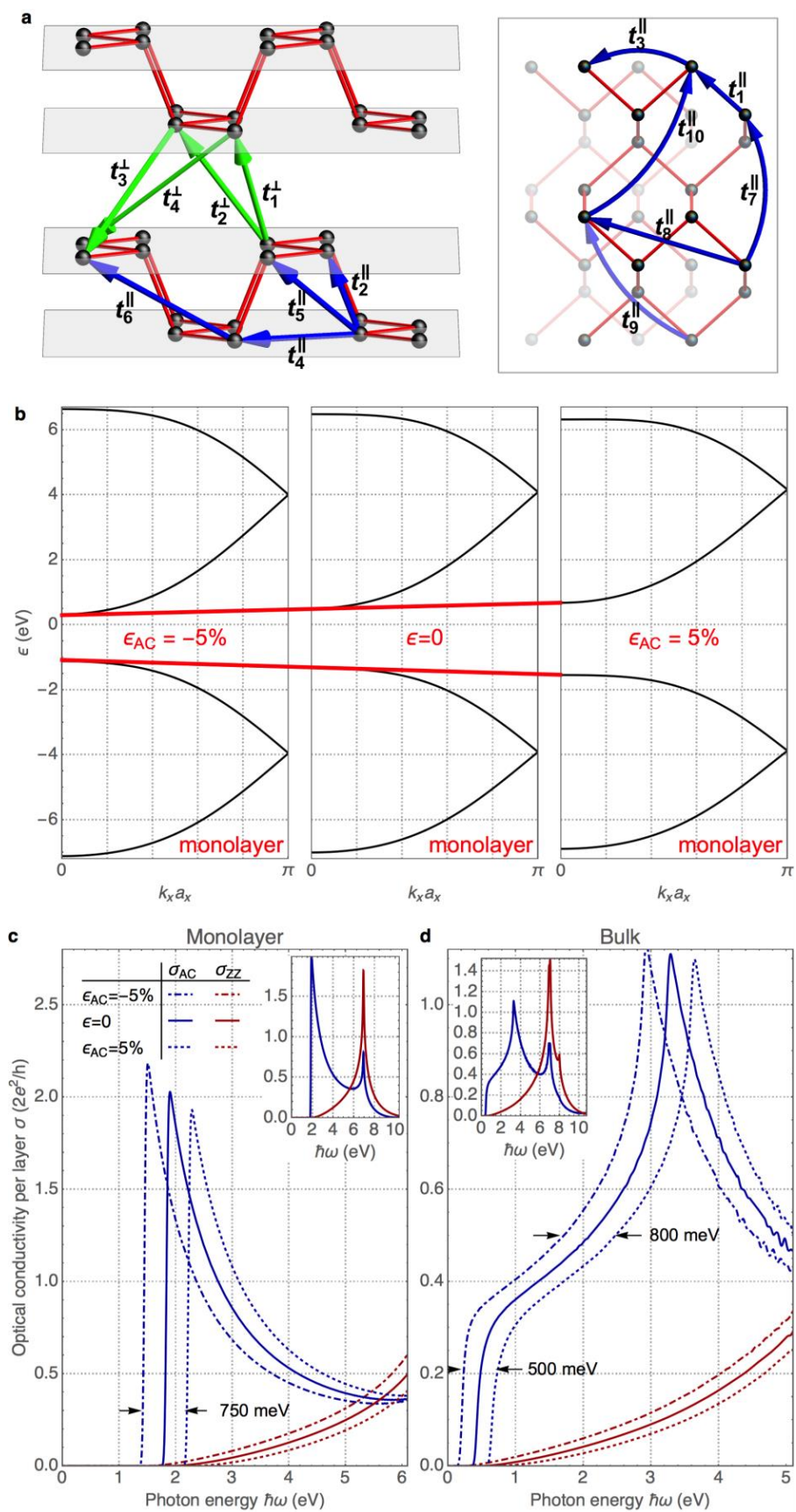

**Figure 2. Black phosphorus under uniform strain: theoretical tight-binding calculations.** **a** Crystal structure of black phosphorus, where the relevant hopping amplitudes considered in the tight-binding model are shown by arrows. **b** Calculated band structure of single layer black phosphorus in the absence of strain (center), under compressive strain (left) and under tensile strain (right). The red lines are guides to the eye that show the evolution of the band edges. **c-d** Optical conductivity of single layer and bulk black phosphorus for different values of applied strain along zigzag ($\sigma_{ZZ}$) and armchair ($\sigma_{AC}$) directions. The full photon energy range is shown in the insets without strains.

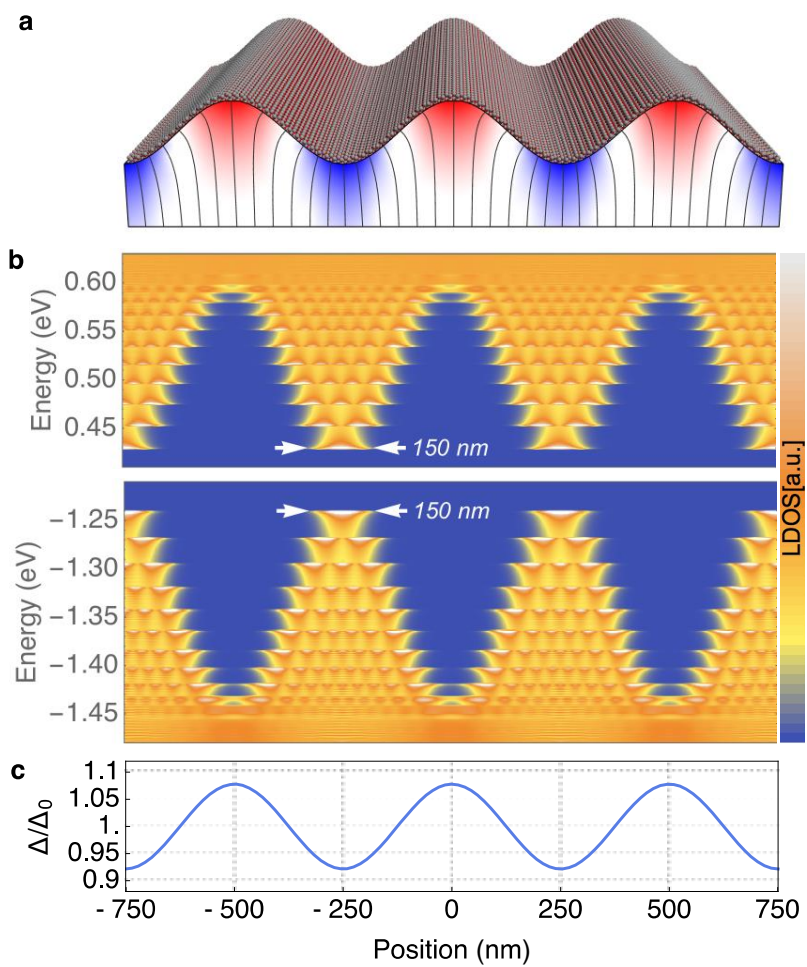

**Figure 3. Black phosphorus under non-uniform strain: theoretical tight-binding calculations.**

**a** Sketch of the rippled black phosphorus monolayer on an elastomeric substrate (white) used to model the experimental system. The cartoon also shows in red (tensile) and blue (compressive), the strength of strain induced on the black phosphorus. The ripple orientation is perpendicular to the armchair axis, and the period is 500 nm. The difference between maximum and minimum strain is 5%. **b** Local density of states, with several quantum-confined channels (fundamental mode ~150 nm wide, spacing 25 meV). **c** Normalized band gap modulation along the ribbon, around ± 1.5% per percentage of uniaxial strain in the monolayer.

ASSOCIATED CONTENT

**Supporting Information**. Supplementary Information includes: black phosphorus thickness determination, determination of the black phosphorus crystal orientation, results on additional black phosphorus samples, generation of the iso-absorption maps, orientation of the ripples with respect to the crystal lattice, details on tight-binding modelling, elasticity and electronic structure analysis, both in the uniform and the non-uniform cases, and including the multilayer black phosphorus. This material is available free of charge via the Internet at http://pubs.acs.org.

AUTHOR INFORMATION

**Corresponding Author**

(Pablo San-Jose) pablo.sanjose@csic.es, (Rafael Roldán) rroldan@icmm.csic.es   and (Andres Castellanos-Gomez) andres.castellanos@imdea.org

**Author Contributions**

The manuscript was written through contributions of all authors. All authors have given approval to the final version of the manuscript.

**Funding Sources**

BBVA Foundation: through the fellowship "I Convocatoria de Ayudas Fundacion BBVA a Investigadores, Innovadores y Creadores Culturales" ("Semiconductores ultradelgados: hacia la optoelectronica flexible"),

MINECO (Spain) through grant FIS2014-57432 and the Ramón y Cajal programme,

The European Research Council, grant 290846

The European Commission under the Graphene Flagship, contract CNECTICT- 604391.

ACKNOWLEDGMENT

The authors acknowledge Herko P. van der Meulen (Autonoma University of Madrid) for providing the tunable monochromatic light source. A.C-G. acknowledges financial support from the BBVA Foundation through the fellowship "I Convocatoria de Ayudas Fundacion BBVA a Investigadores, Innovadores y Creadores Culturales", from the MINECO (Ramón y Cajal 2014 program, RYC-2014-01406) and from the MICINN (MAT2014-58399-JIN). R.R. acknowledges financial support from MINECO (FIS2014-58445-JIN). P.S-J. acknowledges financial support from MINECO (Ramón y Cajal 2013 program, RYC-2013-14645) . The authors also acknowledge support from MINECO (Spain) through grant FIS2014-57432, FIS2011-23713, MAT2014–57915-R, the European Research Council, grant 290846, the European Commission under the Graphene Flagship, contract CNECTICT- 604391 and from the Comunidad de Madrid (MAD2D-CM Program (S2013/MIT-3007), through grant number PIB2010BZ-00512.

REFERENCES

(1) Li, L.; Yu, Y.; Ye, G. J.; Ge, Q.; Ou, X.; Wu, H.; Feng, D.; Chen, X. H.; Zhang, Y. *Nat. Nanotechnol.* **2014**, *9* (5), 372–377.

(2) Koenig, S. P.; Doganov, R. A.; Schmidt, H.; Castro Neto, A. H.; Özyilmaz, B. *Appl. Phys. Lett.* **2014**, *104* (10), 103106.

(3) Wang, X.; Jones, A. M.; Seyler, K. L.; Tran, V.; Jia, Y.; Zhao, H.; Wang, H.; Yang, L.; Xu, X.; Xia, F. *Nat. Nanotechnol.* **2015**, *10* (6), 517–521.

(4) Liu, H.; Neal, A. T.; Zhu, Z.; Luo, Z.; Xu, X.; Tománek, D.; Ye, P. D. *ACS Nano* **2014**, *8* (4), 4033–4041.

(5) Castellanos-Gomez, A.; Vicarelli, L.; Prada, E.; Island, J. O.; Narasimha-Acharya, K. L.; Blanter, S. I.; Groenendijk, D. J.; Buscema, M.; Steele, G. A.; Alvarez, J. V.; Zandbergen, H. W.; Palacios, J. J.; van der Zant, H. S. J. *2D Mater.* **2014**, *1* (2), 025001.

(6) Liu, H.; Du, Y.; Deng, Y.; Ye, P. D. *Chem. Soc. Rev.* **2015**, *44* (9), 2732–2743.

(7) Ling, X.; Wang, H.; Huang, S.; Xia, F.; Dresselhaus, M. S. *Proc. Natl. Acad. Sci.* **2015**, *112* (15), 201416581.

(8) Castellanos-Gomez, A. *J. Phys. Chem. Lett.* **2015**, *6* (21), 4280–4291.

(9) Kim, J.; Baik, S. S.; Ryu, S. H.; Sohn, Y.; Park, S.; Park, B.-G.; Denlinger, J.; Yi, Y.; Choi, H. J.; Kim, K. S. *Science* **2015**, *349* (6249), 723–726.

(10) Gillgren, N.; Wickramaratne, D.; Shi, Y.; Espiritu, T.; Yang, J.; Hu, J.; Wei, J.; Liu, X.; Mao, Z.; Watanabe, K.; Taniguchi, T.; Bockrath, M.; Barlas, Y.; Lake, R. K.; Ning Lau, C. *2D Mater.* **2014**, *2* (1), 011001.

(11) Li, L.; Ye, G. J.; Tran, V.; Fei, R.; Chen, G.; Wang, H.; Wang, J.; Watanabe, K.; Taniguchi, T.; Yang, L.; Chen, X. H.; Zhang, Y. *Nat. Nanotechnol.* **2015**, *10* (7), 608–613.

(12) Chen, X.; Wu, Y.; Wu, Z.; Han, Y.; Xu, S.; Wang, L.; Ye, W.; Han, T.; He, Y.; Cai, Y.; Wang, N. *Nat. Commun.* **2015**, *6*, 7315.

(13) Wei, Q.; Peng, X. *Appl. Phys. Lett.* **2014**, *104* (25), 251915.

(14) Jiang, J.-W.; Park, H. S. *J. Phys. D. Appl. Phys.* **2014**, *47* (38), 385304.

(15) Mehboudi, M.; Utt, K.; Terrones, H.; Harriss, E. O.; Pacheco SanJuan, A. A.; Barraza-Lopez, S. *Proc. Natl. Acad. Sci. U. S. A.* **2015**, *112* (19), 5888–5892.

(16) Roldán, R.; Castellanos-Gomez, A.; Cappelluti, E.; Guinea, F. *J. Phys. Condens. Matter* **2015**, *27* (31), 313201.

(17) Rodin, A. S.; Carvalho, A.; Castro Neto, A. H. *Phys. Rev. Lett.* **2014**, *112* (17), 176801.

(18) Çakır, D.; Sahin, H.; Peeters, F. M. *Phys. Rev. B* **2014**, *90* (20), 205421.

(19) Elahi, M.; Khaliji, K.; Tabatabaei, S. M.; Pourfath, M.; Asgari, R. *Phys. Rev. B* **2015**, *91* (11), 115412.

(20) Conley, H. J.; Wang, B.; Ziegler, J. I.; Haglund, R. F.; Pantelides, S. T.; Bolotin, K. I. *Nano Lett.* **2013**, *13* (8), 3626–3630.

(21) He, K.; Poole, C.; Mak, K. F.; Shan, J. *Nano Lett.* **2013**, *13* (6), 2931–2936.

(22) Hui, Y. Y.; Liu, X.; Jie, W.; Chan, N. Y.; Hao, J.; Hsu, Y.-T.; Li, L.-J.; Guo, W.; Lau, S. P. *ACS Nano* **2013**, *7* (8), 7126–7131.

(23) Castellanos-Gomez, A.; Roldán, R.; Cappelluti, E.; Buscema, M.; Guinea, F.; van der Zant, H. S. J.; Steele, G. A. *Nano Lett.* **2013**, *13* (11), 5361–5366.

(24) Zhu, C. R.; Wang, G.; Liu, B. L.; Marie, X.; Qiao, X. F.; Zhang, X.; Wu, X. X.; Fan, H.; Tan, P. H.; Amand, T.; Urbaszek, B. *Phys. Rev. B* **2013**, *88* (12), 121301.

(25) Plechinger, G.; Castellanos-Gomez, A.; Buscema, M.; van der Zant, H. S. J.; Steele, G. A.; Kuc, A.; Heine, T.; Schüller, C.; Korn, T. *2D Mater.* **2015**, *2* (1), 015006.

(26) Desai, S. B.; Seol, G.; Kang, J. S.; Fang, H.; Battaglia, C.; Kapadia, R.; Ager, J. W.; Guo, J.; Javey, A. *Nano Lett.* **2014**, *14* (8), 4592–4597.

(27) Feng, J.; Qian, X.; Huang, C.-W.; Li, J. *Nat. Photonics* **2012**, *6* (12), 866–872.

(28) Li, H.; Contryman, A. W.; Qian, X.; Ardakani, S. M.; Gong, Y.; Wang, X.; Weisse, J. M.; Lee, C. H.; Zhao, J.; Ajayan, P. M.; Li, J.; Manoharan, H. C.; Zheng, X. *Nat. Commun.* **2015**, *6*, 7381.

(29) Island, J. O.; Steele, G. A.; Zant, H. S. J. van der; Castellanos-Gomez, A. *2D Mater.* **2015**, *2* (1), 011002.

(30) Mei, H.; Landis, C. M.; Huang, R. *Mech. Mater.* **2011**, *43* (11), 627–642.

(31) Qiao, J.; Kong, X.; Hu, Z.-X.; Yang, F.; Ji, W. *Nat. Commun.* **2014**, *5*, 4475.

(32) Yuan, H.; Liu, X.; Afshinmanesh, F.; Li, W.; Xu, G.; Sun, J.; Lian, B.; Curto, A. G.; Ye, G.; Hikita, Y.; Shen, Z.; Zhang, S.-C.; Chen, X.; Brongersma, M.; Hwang, H. Y.; Cui, Y. *Nat. Nanotechnol.* **2015**, *10* (8), 707–713.

(33) Castellanos-Gomez, A.; Quereda, J.; van der Meulen, H. P.; Agraït, N.; Rubio-Bollinger, G. *arXiv* **2015**, *1507.00869*.

(34) Slater, J. C.; Koster, G. F. *Phys. Rev.* **1954**, *94* (6), 1498–1524.

(35) Morita, A. *Appl. Phys. A Solids Surfaces* **1986**, *39* (4), 227–242.

(36) Rudenko, A. N.; Katsnelson, M. I. *Phys. Rev. B* **2014**, *89* (20), 201408.

(37) Rudenko, A. N.; Yuan, S.; Katsnelson, M. I. *Phys. Rev. B* **2015**, *92* (8), 085419.

(38) Suzuura, H.; Ando, T. *Phys. Rev. B* **2002**, *65* (23), 235412.

(39) Pereira, V. M.; Castro Neto, A. H.; Peres, N. M. R. *Phys. Rev. B* **2009**, *80* (4), 045401.

(40) Harrison, W. A. *Elementary Electronic Structure*; World Scientific, 1999.

(41) Castellanos-Gomez, A. *J. Phys. Chem. Lett.* **2015**, *6* (21), 4280–4291.

(42) Low, T.; Rodin, A. S.; Carvalho, A.; Jiang, Y.; Wang, H.; Xia, F.; Neto, A. H. C. *Phys. Rev. B* **2014**, *90*, 075434.

(43) Yuan, S.; Rudenko, A. N.; Katsnelson, M. I. *Phys. Rev. B* **2015**, *91* (11), 115436.

(44) High, A. A.; Leonard, J. R.; Remeika, M.; Butov, L. V; Hanson, M.; Gossard, A. C. *Nano Lett.* **2012**, *12* (5), 2605–2609.

(45) Appalakondaiah, S.; Vaitheeswaran, G.; Lebègue, S.; Christensen, N. E.; Svane, A. *Phys. Rev. B* **2012**, *86* (3), 035105.

# Supporting information:

# Strong modulation of optical properties in black phosphorus through strain-engineered rippling

Jorge Quereda[1], Pablo San-Jose[2,*], Vincenzo Parente[3], Luis Vaquero-Garzon[3], Aday Molina-Mendoza[1,3], Nicolás Agraït[1,3,4], Gabino Rubio-Bollinger[1,4], Francisco Guinea[3], Rafael Roldán[2,3,*], Andres Castellanos-Gomez[3,*]

[1] Dpto. de Física de la Materia Condensada, Universidad Autónoma de Madrid, 28049 Madrid, Spain.
[2] Instituto Madrileño de Estudios Avanzados en Nanociencia (IMDEA-nanociencia), Campus de Cantoblanco, E-18049 Madrid, Spain.
[3] Instituto de Ciencia de Materiales de Madrid, CSIC, Sor Juana Ines de la Cruz 3, 28049 Madrid, Spain.
[4] Condensed Matter Physics Center (IFIMAC), Universidad Autónoma de Madrid, E-28049 Madrid, Spain.

(P.S-J.) pablo.sanjose@csic.es, (R.R) rroldan@icmm.csic.es and (A.C-G.) andres.castellanos@imdea.org

Table of contents:

1) Buckling-induced rippling
2) Black phosphorus thickness determination
3) Determination of the black phosphorus crystal orientation
4) Results on additional black phosphorus samples
5) Photocurrent spectroscopy of unstrained black phosphorus
6) Generation of the iso-absorption maps
7) Orientation of the ripples with respect to the crystal lattice

# 1. Buckling-induced rippling

Figure S1 shows schematically the process followed to fabricate the rippled black phosphorus flakes. The method is based on pre-stretching the PDMS substrate before transferring the black phosphorus flakes. Once the flakes are attached to the PDMS, the substrate is allowed to retract to its original shape. In consequence, the flakes experience an effective compressive force, causing the formation of the ripples.

When a thin elastic film adhered onto a compliant substrate is subjected to uniaxial compressive strain above a certain critical value $\epsilon_c$, it undergoes a rippling process due to the trade-off between the buckling of the thin film and the adhesion between the thin film and the substrate. Specifically, the value of $\epsilon_c$ depends on the plane-strain moduli of the film and the substrate, $\bar{E}_f$ and $\bar{E}_s$ respectively as [S1]

$$\epsilon_c = \frac{1}{4}\left(\frac{3\bar{E}_s}{\bar{E}_f}\right)^{\frac{2}{3}}. \tag{S1}$$

These ripples have a sinusoidal shape given by [S2]

$$d(x) = \frac{\delta}{2}\left[1 + \cos\left(\frac{2\pi x}{\lambda}\right)\right], \tag{S2}$$

where $d(x)$ is the height at a given position x along the ripple. The period of the ripples depends on the ratio of the plane-strain moduli, $\bar{E}_f$ and $\bar{E}_s$, and on the thickness of the thin film, $h$, as

$$\lambda = 2\pi h \left(\frac{\bar{E}_f}{3\bar{E}_s}\right)^{\frac{1}{3}} \tag{S3}$$

And the amplitude of the ripples depends on the ratio between the critical strain value $\epsilon_c$, the applied compression strain $\epsilon$, and the thickness of the thin film, $h$:

$$A = h\sqrt{\frac{\epsilon}{\epsilon_c} - 1} \tag{S4}$$

This sinusoidal ripples are often observed in thin film sample fabrication processes and have has been experimentally reported for smaller samples, including graphene and molybdenum disulfide [S3-S5].

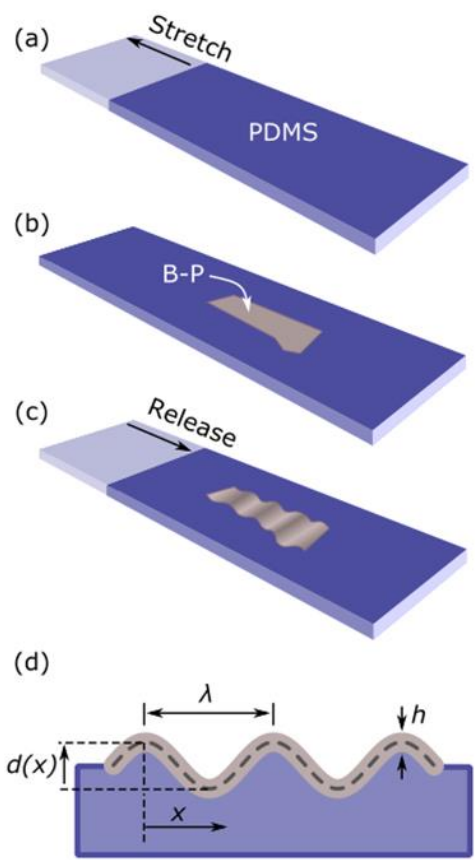

Figure S1 – Process of fabrication of rippled black phosphorus flakes on top of a PDMS substrate. (a) The process starts by laterally stretching a PDMS strip applying a mechanical tensile force. (b) Then, few-layer black phosphorus flakes are transferred to the PDMS from a Nitto tape. (c) Finally, the strip is released, causing its retraction and the black phosphorus flake experiences an effective compressive force, which leads to the formation of ripples. (d) Lateral profile of the rippled flake and PDMS substrate.

## 2. Black phosphorus thickness determination

The thickness determination through atomic force microscopy (AFM) results challenging on top of elastomeric substrates (see the Supporting Information of Ref. [S6]). Moreover, the fast degradation of the exfoliated black phosphorus flakes in ambient conditions makes it necessary to develop a fast and non-destructive method to determine the thickness of the studied black phosphorus flakes.

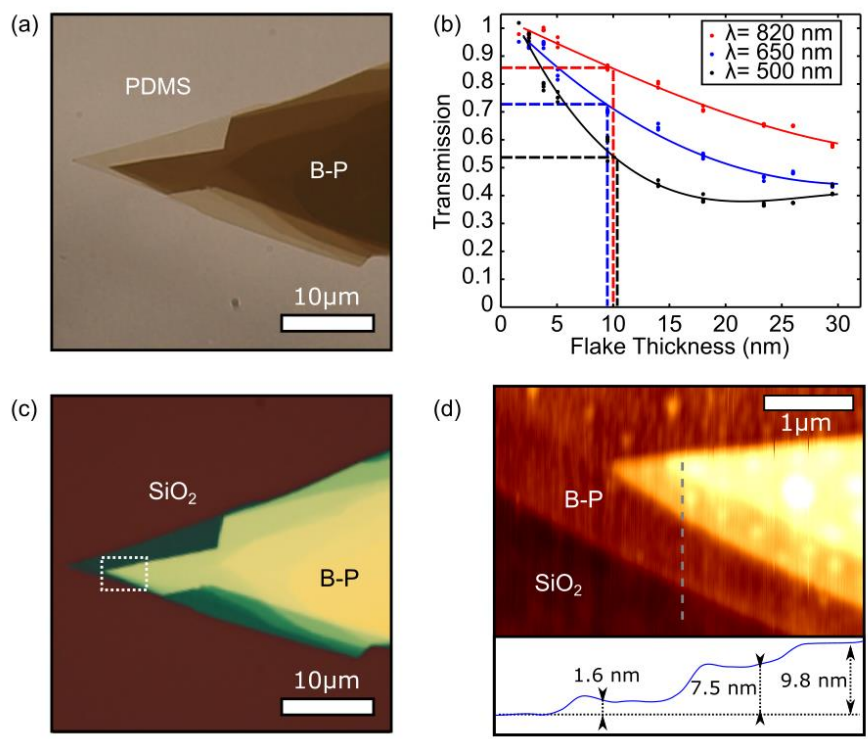

Figure S2. (a) Transmission optical image of a black phosphorus flake with regions of different thicknesses, transferred onto a elastomeric Gelfilm® substrate. (b) Thickness-dependent light transmission spectra of the black phosphorus flake shown in (a) under three different illumination wavelengths. Dots are experimental values of transmission measured in regions of black phosphorus of different thicknesses, measured with an AFM. Solid lines are fits of the experimental data to a 3$^{rd}$ order polynomial. Comparing those fits with the light transmission of the flake shown in the main text it is possible to determine its thickness. We find a thickness value of 10 ± 2 nm. (c) Reflection optical micrograph of the same crystal transferred onto a 300 nm SiO$_2$ / Si substrate to perform AFM measurements. (d) AFM image of the region marked by a rectangle in (c), showing four regions with different thicknesses. Insert: Scan profile along the dashed line marked in (d).

First we measure the light transmission in various regions of black phosphorus flakes with different thicknesses under monochromatic illumination with different wavelengths, selected using optical filters (see Figure S2a). Then we transfer the flakes onto a SiO$_2$/Si substrate, as shown in Figure S2c, using a recently reported transfer technique [S7] and we measure their thickness through contact-mode AFM (Figure S2d).

Figure S2b shows the thickness-dependent light transmission of black phosphorus flakes under three different illumination wavelengths (500 nm, 650 nm and 820 nm), as well as their fit to a 3$^{rd}$ order polynomial function. Comparing those fits with the optical transmission measured at the rippled flakes we are able to determine the thickness of black phosphorus flakes. Using this method we get a thickness of 10 ± 2 nm for the flake shown in Figure 1 of the main text.

## 3. Determination of the black phosphorus crystal orientation

The marked structural anisotropy of black phosphorus is the stem of its anisotropic optical, mechanical and electrical properties. One can exploit its marked linear dichroism, optical absorption that depends on the relative orientation between the materials lattice and incident linearly polarized light, to determine the crystal lattice orientation of a certain black phosphorus flake.

The anisotropic optical properties of black phosphorus are characterized by transmission mode optical microscopy. A linear polarizer is placed between the microscope light source and the condenser lens. Transmission mode images are acquired while the polarizer is rotated in steps of 3°. The transmission is then calculated by normalizing the intensity measured on the black phosphorus flake by the intensity measured on the nearby bare substrate. Figure S3 shows a polar plot with the angular dependence of the transmission of a black phosphorus nanosheet (showed as a water mark inside the polar plot). The polar-plot clearly shows marked linear dichroism: the optical transmission reaches a maximum value when the excitation light is polarized along the zigzag direction.[S8] In this case, the zigzag direction is parallel to a straight edge of the black phosphorus flake.

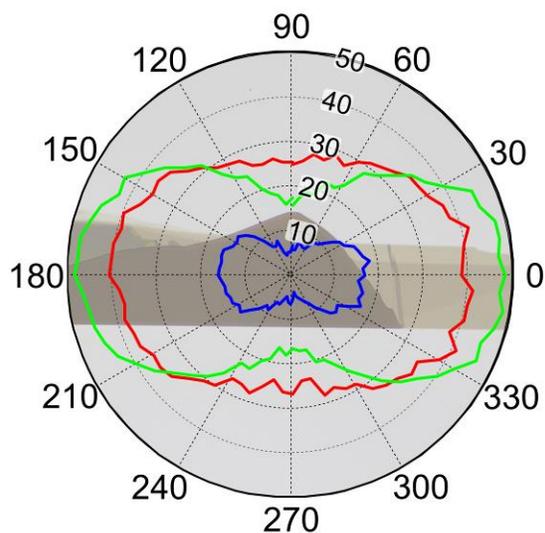

Figure S3 – Angular dependence of the optical transmission, measured on a black phosphorus flake (water marked inside the polar plot) by varying the angle of the linearly polarized illumination. Lines with different colors correspond to the different channels (Red, Green and Blue) of the color CCD camera employed for this measurement.

## 4. Results on additional black phosphorus samples

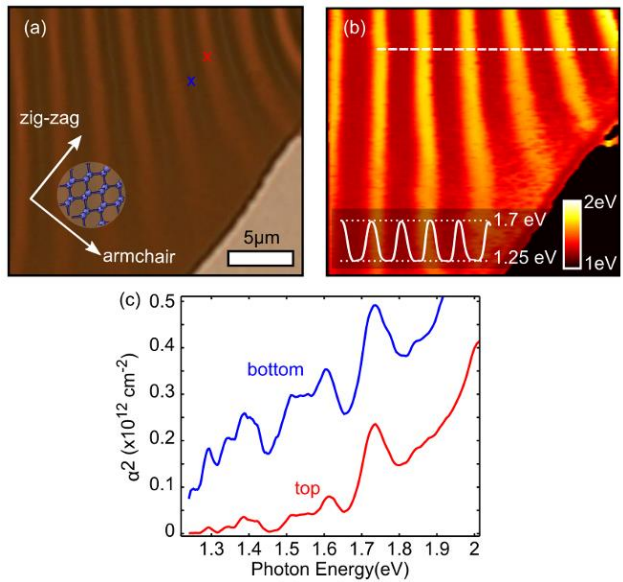

Figure S4 – (a) Transmission optical image of a 30 nm thick rippled black phosphorus flake. Insert: Schematic cartoon indicating the orientation of the crystalline lattice. (b) Iso-absorption map of the same region shown in (a) at $\alpha^2=10^{11}$ cm$^{-2}$. Insert: Iso-absorption energy profile along the dashed line. (c) Squared absorption coefficient as a function of the illumination photon's energy in the two regions marked with 'x' symbols in (a).

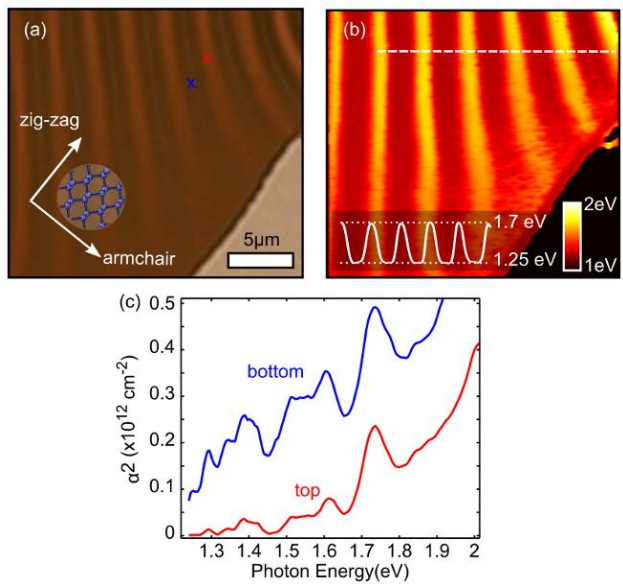

Figure S5 – (a) Transmission optical image of a black phosphorus flake. The region marked by the dashed square has an estimated thickness of 30 nm. Insert: Schematic cartoon indicating the orientation of the crystalline lattice. (b) Iso-absorption map of the region marked in (a) by a dashed square at $\alpha^2=10^{11}$ cm$^{-2}$. Insert: Iso-absorption energy profile along the dashed line. (c) Squared absorption coefficient as a function of the illumination photon's energy in the two regions marked with x symbols in (a).

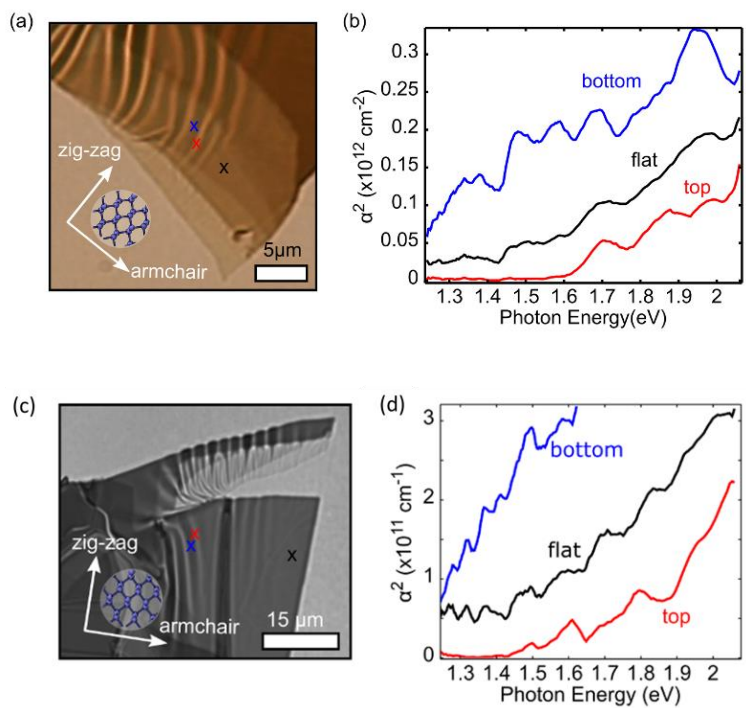

Figure S6 – (a) Transmission optical image of a black phosphorus flake. Insert: Schematic cartoon indicating the orientation of the crystalline lattice. (b) Squared absorption coefficient as a function of the illumination photon's energy in the three regions marked with x symbols in (a), corresponding to a flake thickness of 12 nm. (c) Transmission optical image of a black phosphorus flake. Insert: Schematic cartoon indicating the orientation of the crystalline lattice. (d) Squared absorption coefficient as a function of the illumination photon's energy in the three regions marked with x symbols in (c), corresponding to a flake thickness of 25 nm.

## 5. Photocurrent spectroscopy of unstrained black phosphorus

The absorption spectra in Figure 1b drops (in the case of the topmost part of the ripples even below the resolution limit of our experimental setup) at 1.4-1.8 eV, In order to determine whether the absorption drop observed in the 1.4-1.8 eV range is related to the fundamental absorption edge (transitions between the top of the valence band and the bottom of the conduction band) or is due to a higher energy optical transition we have fabricated a ~10 nm thick black phosphorus photodetector and we studied its photocurrent (which is approximately proportional to the absorption) as a function of the illumination wavelength. This approach allows us to overcome the limitation of our detector.

The black phosphorus photodetector is fabricated by deterministic transfer of the mechanically exfoliated material on pre-patterned Au electrodes on $SiO_2/Si^+$.[S7] Briefly, the bulk material is exfoliated with an adhesive tape (Nitto tape), obtaining the exfoliated source material which is subsequently exfoliated with a polydimethylsiloxane (PDMS, Gelfilm from Gelpak) stamp. Several flakes with different thicknesses remain adhere to the stamp and their thickness can identified by optical microscopy in transmission mode. The desired flake is then deterministically transferred on the substrate, consisting of two Ti/Au (5 nm / 50 nm) electrodes, separated 10 µm, which have been previously patterned by electron-beam induced shadow-mask evaporation on a $SiO_2$ layer (285 nm) thermally grown on a highly p-doped Si substrate. The sample is then annealed at 100 ºC in vacuum (P ~ 1 mbar) during 20 minutes to improve the electrical contact between the flake and the electrodes. In Figure S7a we can see an optical microscopy image of the fabricated device.

The optoelectronic characterization is performed in a homemade probe station operating in ambient conditions (air atmosphere), right after sample preparation to avoid degradation of the material.[S9] Figure S7b shows an artistic representation of the measurement: a light spot with a known diameter, wavelength (provided by a LED source, Thorlabs) and power (previously measured) is used to illuminate the material with modulated intensity at 0.25 Hz. The drain-source current is measured as a function of time while the LED is switched on and off, allowing to determine the photocurrent ($I_{ph}$), defined as the difference of the current upon illumination and in drak conditions, of the device (Figure S7c). From this measurement with obtain photocurrent generation in a wide range of wavelengths from the ultraviolet (UV) up to the near-infrared (NIR) region of the spectrum, in good agreement with previously reported experiments.[S10] We can also calculate the responsivity ($R$) of the device as a function of the light wavelength, defined as the ratio between the generated photocurrent and the light power: $R = I_{ph}/P$. The responsivity is usually calculated using the light effective power ($P_{eff}$) instead of the total power, giving a more accurate number, which is defined as: $P_{eff} = P \cdot A_{dev}/A_{spot}$, where $A_{dev}$ is the area of the flake lying in between the electrodes and $A_{spot}$ is the area of the light spot. The calculated responsivity is shown in Figure S7d, where we see a maximum value of ~ 90 mA·W$^{-1}$ upon illumination with light wavelength of 1500 nm, in good agreement with previously reported values.[S11]

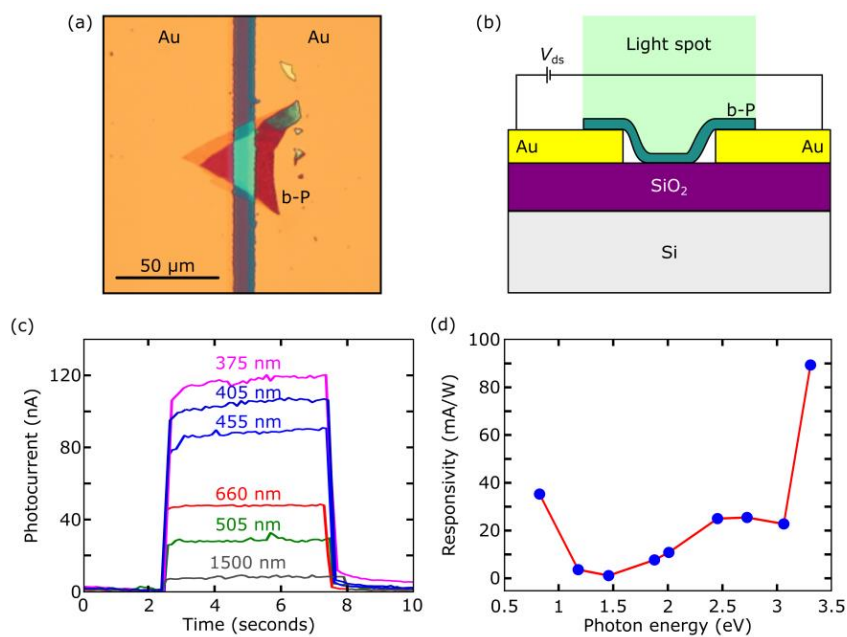

Figure S7 (a) Optical microscopy image of the photodetector based on a black-phosphorus flake. (b) Artistic representation of the photocurrent measurement of the photodetector based on a black-phosphorous flake. (c) Photocurrent (difference between the current upon illumination and in dark conditions) time response of the black-phosphorus based photodetector for different wavelengths. (d) Responsivity as a function of the LED photon energy.

## 6. Generation of iso-absorption maps

In order to generate the iso-absorption maps we first acquire hyperspectral images of the rippled flakes as explained in detail in Ref. [S12]. Then, we calculate the wavelength-dependent absorption coefficient, α, for each location on the sample by the using:

$$\alpha = - \mathrm{Ln}(T) / d$$

where $T$ is the light transmission (*i.e.* the ratio between the light intensity transmitted by the flake and the light intensity transmitted by the substrate) and $d$ is the flake's thickness.

Finally, at each sample location the excitation energy value at which the $\alpha^2$ value matches a certain cut-off value (iso-absorption energy) is extracted and it is represented in a color map form.

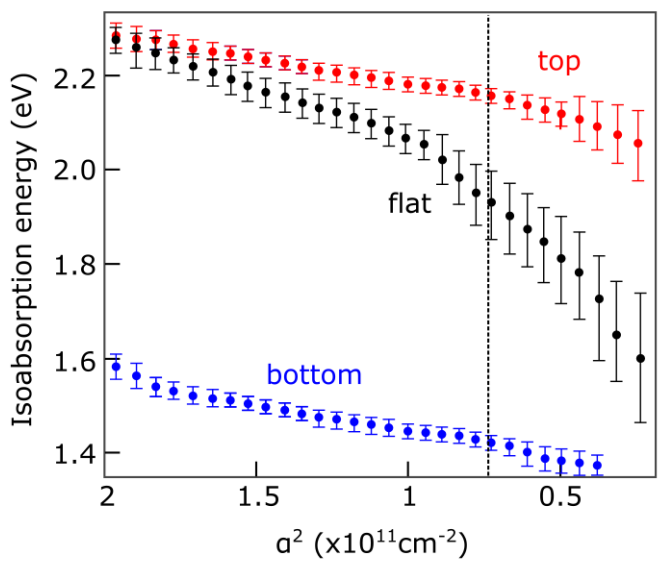

Figure S8: Iso-absorption energy values obtained from the dataset displayed in Figure 1c (main text) for different $\alpha^2$ threshold values. Note that for lower and lower threshold values the iso-absorption energy becomes better and better estimate of the band gap energy. For low $\alpha^2$ threshold values, on the other hand, the uncertainty in the determination of the isoabsorption energy increases. Thus we have employed $7.5 \cdot 10^{-11}$ cm$^{-2}$, value where the uncertainty starts to increase more pronounced, to prepare the Figure 1c of the main text.

## 7. Orientation of the ripples with respect to the crystal lattice

During this work we found out that the ripples on the black phosphorus samples tend to be aligned preferentially parallel to the zigzag direction. Figure S9 shows examples of rippled black phosphorus flakes whose crystal lattice orientation has been determined by measuring their linear dichroism on a flat region of the flake. Figure S9h shows a histogram with the angles formed between the ripples and the zigzag direction which illustrates how the ripples typically occur around the zigzag direction. This is in agreement with recent theoretical works that predict that the zigzag direction is about 4 times stiffer than the armchair direction.

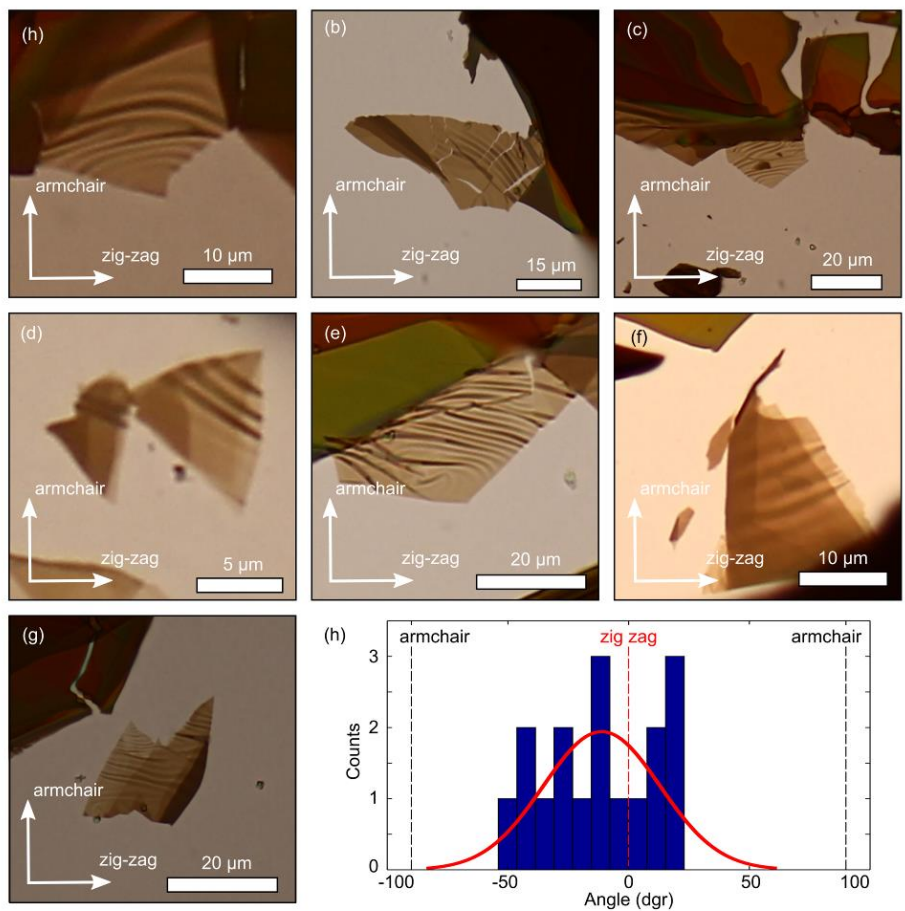

Figure S9 – (a-g) Optical images of various thin black phosphorus crystals with ripples, showing the orientation of the zigzag and armchair crystalline directions (h) Histogram showing the orientation of valleys and peaks from different flakes. The solid red line is a Gaussian fit of the histogram. The valleys and peaks tend to align with the zigzag direction of the crystalline lattice.

## 8. Tight Biding Model

Black phosphorus lattice (Figure S10a) is rectangular with lattice parameters $a_1 = 3.3$ Å and $a_2 = 4.64$ Å, each unit cell containing 4 atoms in a puckered structure. By fitting with *Ab Initio*

$$H = \sum_{\langle\langle i,j \rangle\rangle} t_{ij}^{\parallel} c_i^{\dagger} c_j + \sum_{\langle\langle i,j \rangle\rangle} t_{ij}^{\perp} c_i^{\dagger} c_j$$

calculations, it has been shown in Ref. [S13] that the electronic band structure of black phosphorus can be well described by a suitable tight-binding model

(S5)

where the $t_{ij}^{\parallel}$ are intra-layer hoppings and $t_{ij}^{\perp}$ are the inter-layer hoppings. The relevant hopping amplitudes are shown in Figure S10(a) and their corresponding values are given in Table S1. Band structure for black phosphorus monolayer, obtained from (S5) putting $t_{ij}^{\perp} = 0$ is shown in Figure S10(b). In Figure S10(c) we plot the energy bands for a sample with 18 layers, which is the typical thickness of the samples studied in our experiments.

| No. | $t^{\parallel}(eV)$ | No. | $t^{\parallel}(eV)$ | No. | $t^{\perp}(eV)$ |
|---|---|---|---|---|---|
| 1 | -1.486 | 6 | 0.186 | 1 | 0.524 |
| 2 | 3.729 | 7 | -0.063 | 2 | 0.180 |
| 3 | -0.252 | 8 | 0.101 | 3 | -0.123 |
| 4 | -0.071 | 9 | -0.042 | 4 | -0.168 |
| 5 | -0.019 | 10 | 0.073 | | |

**Table S1**. Values of the tight-binding parameters for the in-plane and out-of-plane hopping amplitudes [S13].

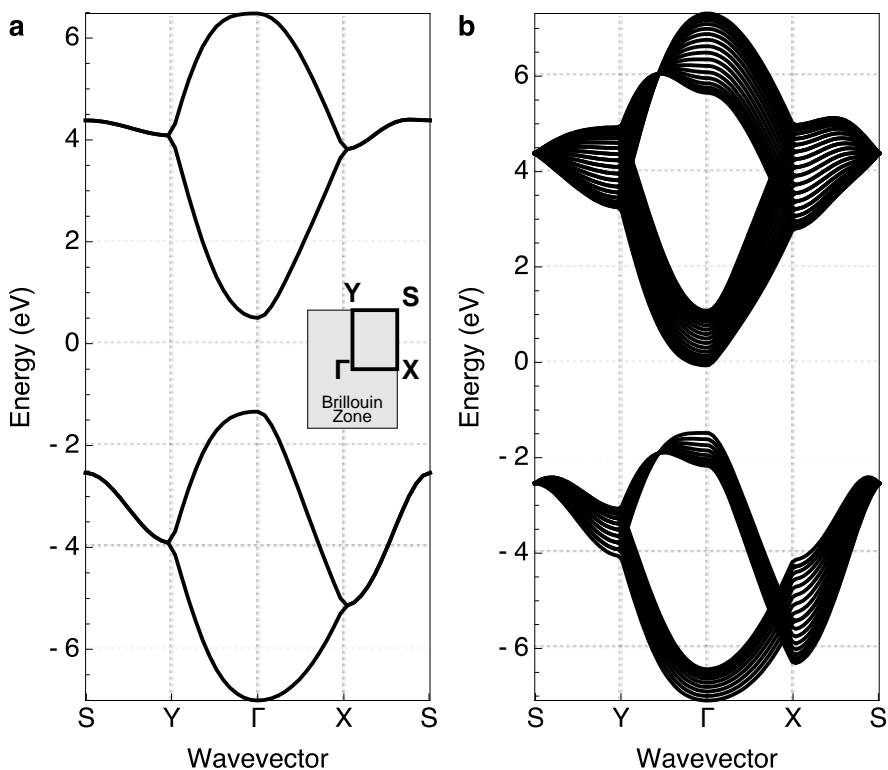

Figure S10: (a) Energy bands for black phosphorus monolayer obtained from the tight-binding model. The Brillouin zone is shown in the inset. (b) Band structure for a sample with 18 layers.

## 9. Strain Effects on Band Gap: the uniform case

As explain in Methods, within the Slater-Koster scheme the hopping amplitudes are functions of the interatomic distances. When the sample is subjected to an external strain, such distances are modified according to the corresponding strain tensor, and the new atomic positions determine the modification to the hopping amplitudes. The change of the most relevant hopping terms with strain is shown in Figure S10 for uniaxial strains along the two principal directions of the lattice. As discussed in the main text, the gap in the monolayer is controlled to leading order by the combination $\Delta \approx \left(4t_1^{\parallel} + 2t_2^{\parallel}\right)$. Since the two hopping amplitudes have different sign (see Table S1), and $t_2^{\parallel}$ is much less sensitive to external strain than $t_1^{\parallel}$, as inferred from Figure S11, the effect of strain on the gap size is much more important in phosphorene than in other 2D crystals as $MoS_2$, where the size of the gap mainly depends on the crystal fields of the orbitals of the metal atoms, which are almost independent of strain.

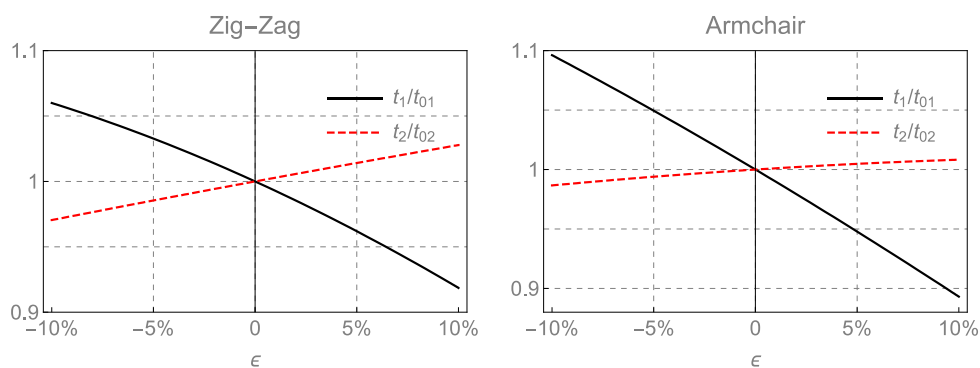

Figure S11: Hopping amplitudes $t_1^\parallel$ and $t_2^\parallel$ as a function of strain, normalized to their corresponding value in the undistorted lattice ($t_{01}^\parallel$ and $t_{02}^\parallel$ respectively). Panel (a) corresponds to uniaxial strain along the zigzag direction, and (b) along the armchair direction, which has a more pronounced dependence.

In Figure S12 we present the modulation of the gap due to shear strain. The gap variation due to shear strain does not depend on its direction (armchair or zig-zag).

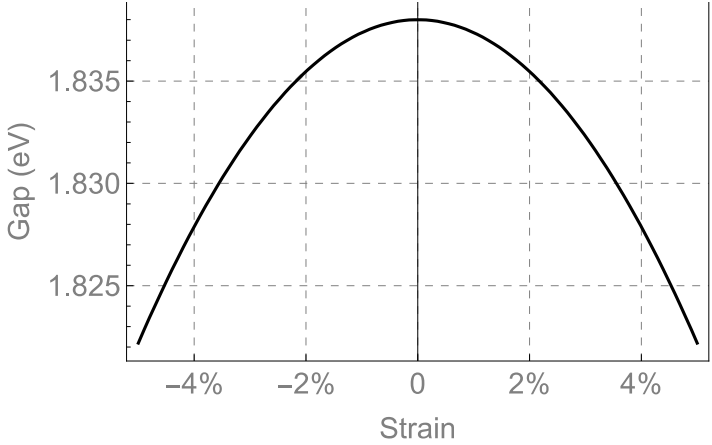

Figure S12: Band gap renormalization due to shear strain in a black phosphorus monolayer.

# 10. Optics in Black Phosphorus

In order to match with experimental results about optical absorption, we have computed the (real part of the) optical conductivity as a function of photon energy using the Kubo formula,

$$\sigma_{ii}(\omega) = \frac{e^2}{h}\frac{1}{A\omega}\sum_{mn,\vec{k}}(f(E_n) - f(E_m))\left|\left\langle\Psi_n(\vec{k})|v_i|\Psi_m(\vec{k})\right\rangle\right|^2 \delta[\hbar\omega - (E_n(\vec{k}) - E_m(\vec{k}))]. \tag{S6}$$

In (S6) $A$ is the area of the unit cell, $\Psi_n(\vec{k})$ is the eigenstate relative to energy $E_n$, $f(E_n) = 1/(1 + e^{\beta E_n})$ is the Fermi distribution, considering that the Fermi level is taken inside the gap, and the velocity operators.

Linear polarized light gives access to anisotropic features of $\sigma(\omega)$. It has already been predicted [S14,S15] that conductivity along the armchair direction $\sigma_{AC}(\omega)$ is an order of magnitude larger than along zig-zag direction $\sigma_{ZZ}(\omega)$. In Figure 2c and 2d of the main text these two component, normalized to the conductance quantum $\sigma_0 = 2e^2/h$, are plotted as a function of the photon energy $\hbar\omega$. In Figure S13 we present the optical conductivity as a function of the polarization angle $\theta$ of incident light

$$\sigma(\theta) = \sigma_{xx} \cos^2 \theta + \sigma_{yy} \sin^2 \theta.$$

As reported in References [S14] and [S15] the 8-like shape can be smoothed into an ellipsoidal shape by the presence of disorder, such as vacancies or ad-atoms. This smoothing is compatible with the experimental curves for optical absorption shown in Section 2 of this Supplementary Information.

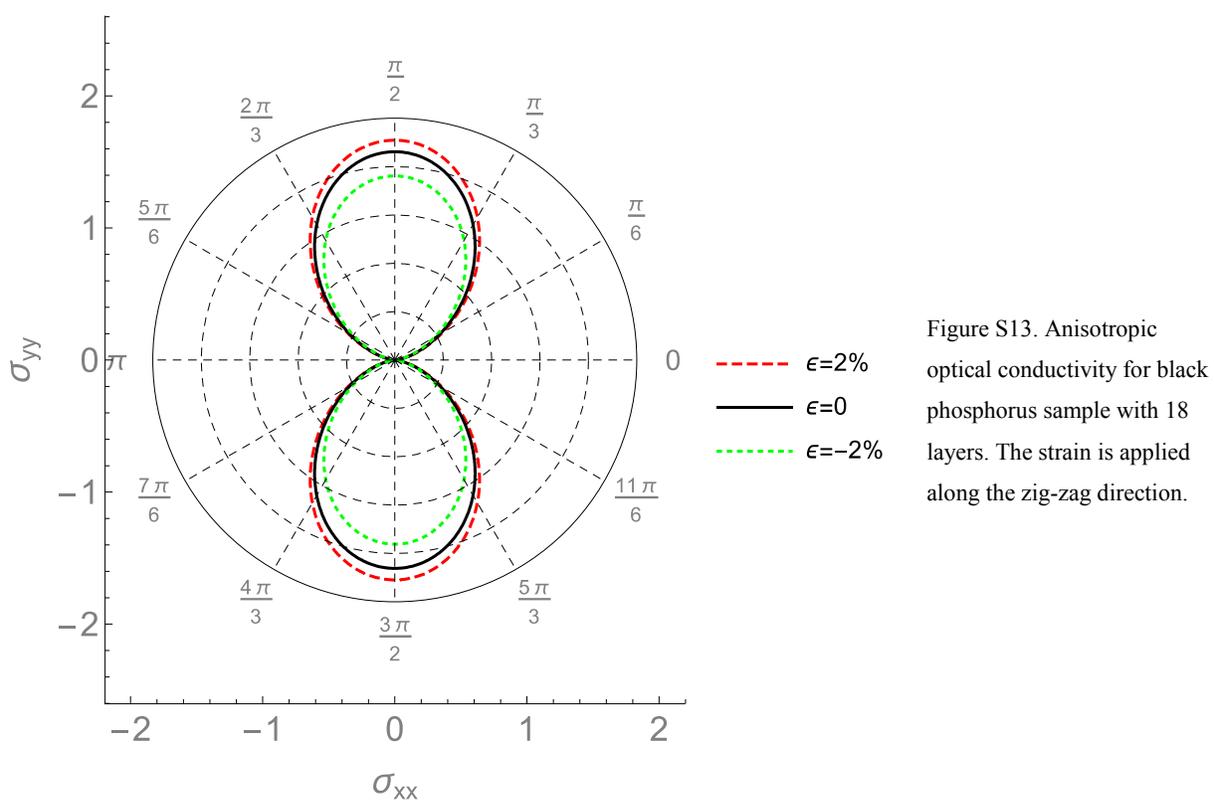

Figure S13. Anisotropic optical conductivity for black phosphorus sample with 18 layers. The strain is applied along the zig-zag direction.

## 11. Non-uniform Strain in phosphorene nanoribbons

Motivated by the dependence of gap on strain highlighted in the previous sections, here the effects of non-uniform uniaxial stress will be analyzed. We model the experimental situation with a sinusoidal uniaxial strain profile along the armchair direction

$$\varepsilon_{AC} = \epsilon(y) \begin{pmatrix} -\nu_{xy} & 0 & 0 \\ 0 & 1 & 0 \\ 0 & 0 & -\nu_3 \end{pmatrix}$$

where $\epsilon(y) = \frac{\epsilon_0}{2}\cos(2\pi y/L)$. We now consider the local density of states (LDOS) in the presence of this non-uniform strain. The LDOS is defined as

$$\rho(E,\vec{r}) = \sum_{n,\vec{k}} \left|\langle \vec{r}|\Psi_n(\vec{k})\rangle\right|^2 \delta\left(E - E_n(\vec{k})\right)$$

where $E_n(\vec{k})$ and $|\Psi_n(\vec{k})\rangle$ are the eigenvalues and eigenstates of the Bloch Hamiltonian $H(\vec{k})$ defined on the ripple unit cell, of size $L_x \times L_y$. We consider ripples perpendicular to the armchair direction, of period $L_y$. The above equation is evaluated by discretizing the Brillouin zone and approximating the δ functions by narrow normalized Gaussians.

The LDOS $\rho(E,y)$ was computed for an infinite rippled monolayer, with a period $L_y = 500$nm and a strain variation $\epsilon_0 = 5\%$. The results are shown in Figure 3b of the main text. The modulation of the gap produced by the sinusoidal strain produces quantum confinement of carriers into discrete 1D channels. The lowest modes in both conduction and valence bands are localized within ~150 nm of the ripple valleys, and are hence only disperse along the ripples ($y$ direction). The subband spacing in this case is around 25 meV, i.e. around room temperature.

We have furthermore analysed the LDOS for ripples perpendicular to the zig-zag edge, for a ripple period of 250 nm, see Fig. S14. The gap modulation is similar, although somewhat smaller, than for the armchair case. This similarity is expected, as the gap-modulation mechanism relies on the change of $t_{2\parallel}/t_{1\parallel}$ with uniaxial strain, which is roughly equivalent in all directions.

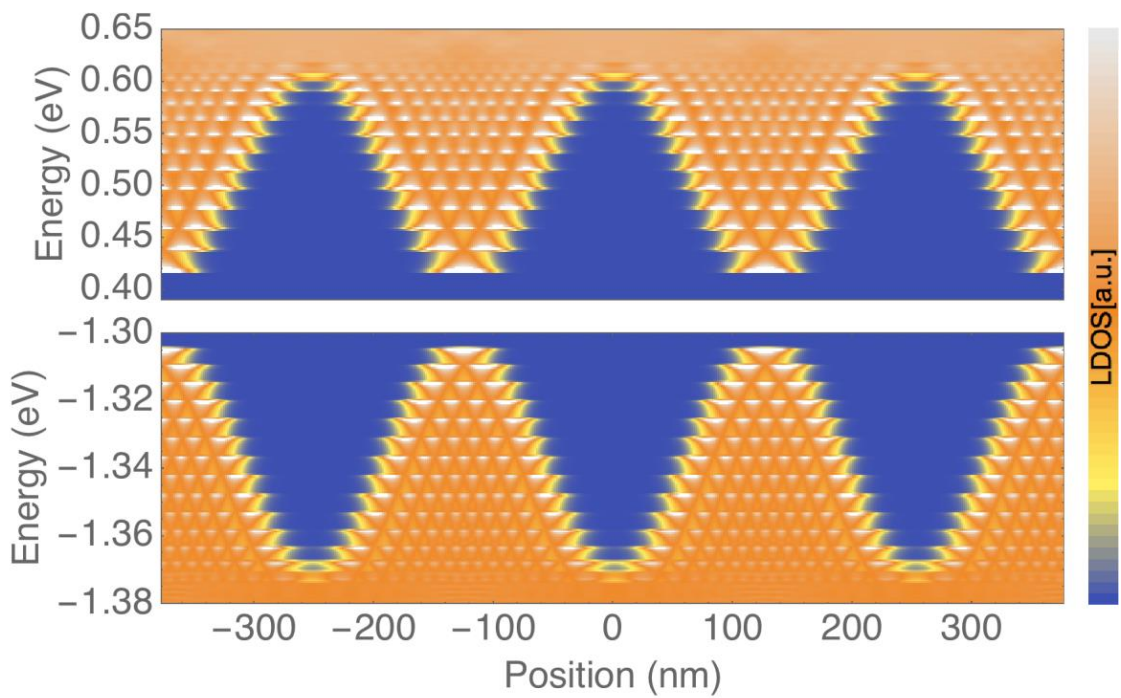

Figure S14. Local density of states, with several quantum-confined channels for zig-zag ripples with a 250 nm period.

The experimental sample has an estimated thickness of about 18 layers. We now show that the above qualitative picture of quantum confinement in the monolayer carries over to the realistic multilayer case. We first consider a bilayer, with a period $L_y = 250$ nm, and the same value for the strain $\epsilon_0 = 5\%$. The LDOS is shown in Figure S15(a,b) for the conduction and valence bands, respectively. We note the same pattern of discrete 1D channels developing in this case, since the period $L_y$, albeit shorter than in the previous simulation, is still larger than the quantum confinement length ~100 nm. The subband spacing is ~50 meV.

An alternative way to picture the effect of quantum confinement is to plot the spatial density of the quantum wire eigenmodes for zero momentum along the ripples $k_x = 0$ (i.e. $|\langle y|\Psi_n(k_y)\rangle|^2$). This is shown in Figures S15(c,d), together with the dispersion of the subbands (dashed curves) along the ripples (i.e. $E_n(k_y)$). We see that different modes are characterized by an increasing number of nodes and larger spatial width. The width eventually reaches the lattice period when the mode has an energy comparable to the local gap at the summit (fifth mode –blue- in Figure S15(c,d)), at which point the mode ceases to be confined and disperses along both $x$ and $y$ directions.

For a rippled sample with 18 layers, the mode bandstructure is considerably denser, as can be seen in Figure S15(e,f). However, both the confinement length and subband spacing of the lowest mode remains comparable to those of the monolayer, at ~100 nm and ~20 meV respectively.

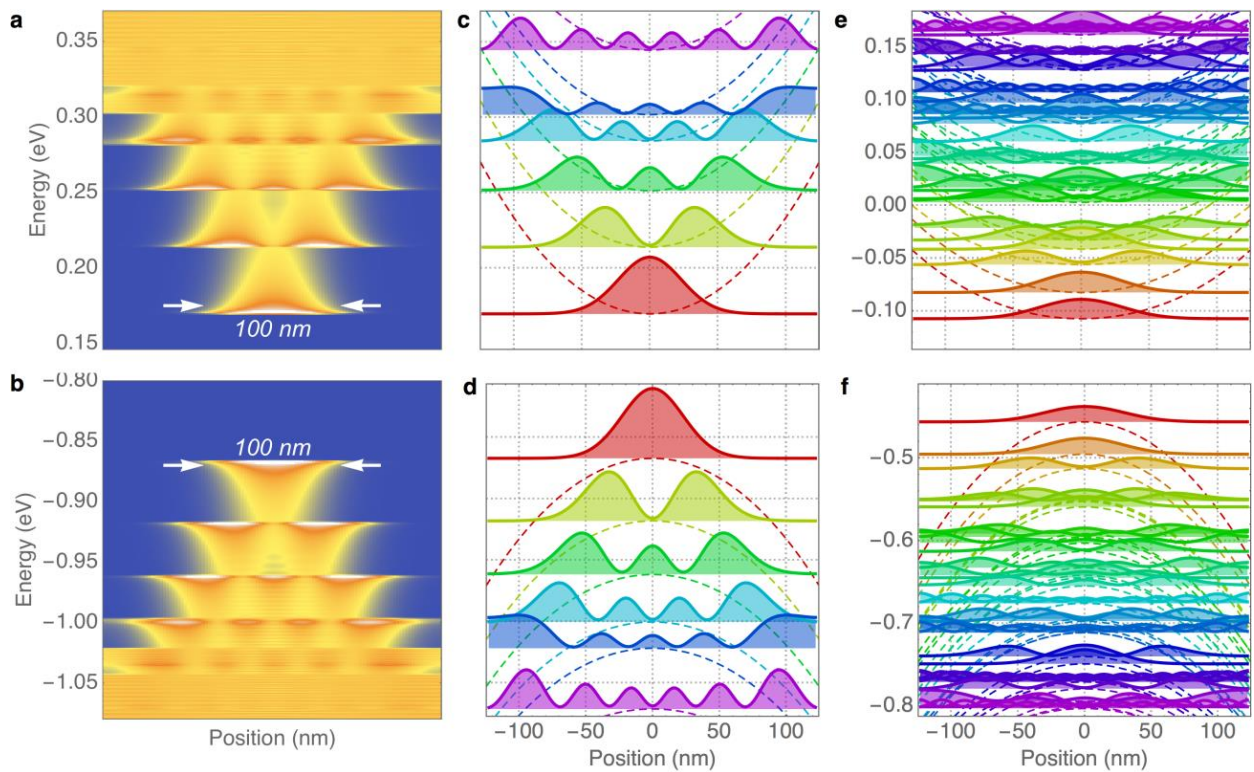

Figure S15. Local density of states and bandstructure of a black phosphorus multilayer. (a,b) Local density of states for a rippled bilayer with ripple period $L_y = 250$nm, and maximum strain modulation $\epsilon_0 = 5\%$. (c,d) Wavefunction probability density of the quantum-confined 1D modes as a function of position perpendicular to the ripple. Dashed lines show their dispersion along the ripple. Only energies within the gap modulation, for which confinement occurs, are shown. (e,f) Same as (c,d) for an 18 layer rippled sample, exhibiting similar spatial width and subband spacing of the fundamental modes (red).

## Supporting information references

S1. Mei, Haixia; Landis, Chad M.; Huang, Rui. *Mechanics of Materials* 2011 **43**(11) p. 627-642.

S2. Vella, D.; Bico, J.; Boudaoud, A.; Roman, B.; Reis, P. M. The Macroscopic Delamination of Thin Films from Elastic Substrates. *Proc. Natl. Acad. Sci. U. S. A.* 2009 **106**, 10901–10906.

S3. Reyes-Martinez, M. a.; Ramasubramaniam, A.; Briseno, A. L.; Crosby, A. J. The Intrinsic Mechanical Properties of Rubrene Single Crystals. *Adv. Mater.* 2012 **24**, 5548–5552.

S4. Wang, Y.; Yang, R.; Shi, Z.; Zhang, L.; Shi, D.; Wang, E.; Zhang, G. Super-Elastic Graphene Ripples for Flexible Strain Sensors. *ACS Nano,* 2011 **5**, 3645–3650.S5. Brennan, Christopher J.; Nguyen, Jessica; Yu, Edward T.; Lu, Nanshu *Advanced Materials Interfaces,* 2015 **2** (16)

S6. Castellanos-Gomez, A. et al. *Local strain engineering in atomically thin MoS2*. Nano Lett. **13**, 5361–6 (2013).

S7. Castellanos-Gomez, A., et al., *Deterministic transfer of two-dimensional materials by all-dry viscoelastic stamping.* 2D Materials, 2014. **1**(1): p. 011002.

S8. Qiao, J., Kong, X., Hu, Z.-X., Yang, F. & Ji, W. High-mobility transport anisotropy and linear dichroism in few-layer black phosphorus. *Nat. Commun.* **5,** 4475 (2014).

S9. Island, J. O., Steele, G. A., Zant, H. S. J. van der & Castellanos-Gomez, A. Environmental instability of few-layer black phosphorus. *2D Mater.* **2,** 011002 (2015).

S10. Engel, M., Steiner, M., and Avouris, P.. Black phosphorus photodetector for multispectral, high-resolution imaging. *Nano letters* **14**, 6414-6417 (2014).

S11. Youngblood, Nathan, et al. Waveguide-integrated black phosphorus photodetector with high responsivity and low dark current. *Nature Photonics* (2015).

S12. Castellanos-Gomez, A., Quereda, J., van der Meulen, H. P., Agraït, N. & Rubio-Bollinger, G. Spatially resolved optical absorption spectroscopy of single- and few-layer MoS2 by hyperspectral imaging. (2015). at <http://arxiv.org/abs/1507.00869>

S13. A. N. Rudenko and M. I. Katsnelson. *Quasiparticle band structure and tight-binding model for single- and bilayer black phosphorus*, Phys. Rev. B **89**, 201408 (2014).

S14. S. Yuan, A. N. Rudenko, and M. I. Katsnelson, *Transport and Optical Properties of Single- and Bilayer Black Phosphorus with Defects*, Phys. Rev. B **91**, 115436 (2015)

S15. T. Low, A. S. Rodin, A. Carvalho, Y. Jiang, H. Wang, F. Xia, and A. H. Castro Neto, *Tunable optical properties of multilayer black phosphorus thin films*, Phys. Rev. B, **90**, 75434 (2014)